\documentclass{aastex}
\usepackage{spr-astr-addons}
\usepackage{url}

\RequirePackage{color}

\begin{document}

\title{On the fraction of particles involved in magneto-centrifugally generated ultra-high energy electrons in the Crab pulsar}




\author{Osmanov, Z.N.\altaffilmark{1,2}}
\affil{School of Physics, Free University of Tbilisi, 0183, Tbilisi,
Georgia email: z.osmanov@freeuni.edu.ge}
\affil{E. Kharadze Georgian National Astrophysical Observatory, Abastumani 0301, Georgia}
\author{Mahajan, S.M.\altaffilmark{3}}
\affil{nstitute for Fusion Studies, The University of Texas at
Austin, Austin, TX 78712, USA}

\begin{abstract}
The earthward journey of ultra high energy electrons ($\sim 600$ TeV)
produced in the Pulsar atmosphere by Landau damping of magneto-centrifugally
excited Langmuir waves (drawing energy form the rotational slowdown)
on primary electrons, is charted. It is shown, that just as they
escape the light cylinder zone, the ultra-high energy particles, interacting
with the medium of the Crab nebula, rapidly loose their energy via
the quantum synchrotron process, producing highly energetic gamma rays ~ $\sim
0.6$PeV. Interacting with the cosmic background radiation in the interstellar medium,
only a tiny fraction of these ultra high energy photons  (via the $\gamma\gamma$ channel) are, then
transformed into electron-positron pairs. Detected flux of these photons
imposes an upper limit on the fraction ($4\times 10^{-7}$) of the magnetospheric particles involved in the process of generation of ultra-high energy photons (up to $600$ TeV).
 
\end{abstract}



\section{Introduction }

A theoretical framework  for particle acceleration, driven by the
rotational slow down of a pulsar, has been proposed in which the
star rotational energy is channeled to the particles in a two step
process - the building up of Langmuir waves (via a two-stream
instability) in the electron-positron plasma in the pulsar
magnetosphere; and the  damping of Langmuir waves on the fast
electrons accelerating them to even higher energies up to 1 PeV
\citep{galaxy,screp,incr}.

It was also surmised that the high energy particles produced, for
example, in the  magnetosphere of the Crab pulsar ($B_{st}\sim
7\times 10^{12}$G), will,  directly or indirectly, manage to reach
detectors on earth with a considerable remnant energy.

In this paper, we tarck the possible journey of such highly
energetic particles as they pass through the nebular, and then the
galactic medium. After an interesting series of mechanisms -
including quantum synchrotron process and conversion to high energy gamma rays, we will  demonstrate that $0.6$ PeV energy photons from the Crab nebula might reach an earth-based detector. By taking into account the observed integral limit on gamma-ray emission, we also estimate the upper limit on magnetospheric particles participating in the energization process 

Before working out the  details of the dynamics of the PeV
electrons, we would summarize the detailed dynamics that led to the
creation of such energetic particles. We have concentrated on the
Crab pulsar whose rotational slowdown is the fundamental energy
source that is channelized for particle acceleration.

The crab pulsar - a rapidly rotating neutron star is a most interesting
astrophysical object characterized by relatively high angular
velocities of rotation. Belonging to the class known as millisecond
pulsars, its rotation period $P\approx 0.0332$s, is decreasing at
the rate $\dot{P}\equiv dP/dt\approx 4.21\times 10^{-13}$ss$^{-1}$.
The corresponding energy released per second, as a result of the
slowdown, is enormous: $\dot{W}\approx I\Omega d\Omega/dt\approx
5\times 10^{38}$erg/s, where $I= 2/5 MR_{st}^2$ is the moment of
inertia of the Crab pulsar, $M\sim (1.5-2.5)\times M_{\odot}$ is its
mass, $M_{\odot}\approx 2\times 10^{33}$g is the solar mass,
$R_{st}\approx 10$km is the neutron star's radius and $\Omega\equiv
2\pi/P$ is the pulsar's angular velocity of rotation. It is worth
noting that only a tiny fraction of the total energy goes through
the pulsar channel, approximately $40\%$ is radiated by the nebula
and the rest is lost in unknown ways.

In the framework of the standard pulsar model \citep{deutsh,gj}, the
magnetosphere is fed by the electrons uprooted from a neutron star's
surface by means of the longitudinal component (along the magnetic
field lines) of the electrostatic field. In due course of time these
particles accelerate and start emitting radiation because of the
nonzero curvature of field lines. It is strongly believed that the
photons inevitably reach the threshold energy, $2mc^2$ ($m$ is the
electron's mass and $c$ is the speed of light) when they transit to
the pair creation channel $\gamma + {\bf B}\rightarrow
e^{+}+e^{-}+\gamma'$ leading to the cascading mechanism; the pair production 
continues until the magnetospheric plasma electron-positron ($e^+e^-$) plasma screens out the initial electrostatic field. We have
discussed in a previous paper that the acceleration inside the gap
(augmented by several modifications) is  quite inadequate, by itself,  to explain
the observed high energy emission from the Crab pulsar
\citep{screp}.

Our model of particle acceleration to high energies, the Langmuir Landau --- (LLCD) mechanism \citep{galaxy,screp,incr},
is a two stage process: 1) the time dependent centrifugal force excites
growing  electrostatic waves in the  electron-positron plasma
permeating the pulsar magnetosphere , 2) these waves in
turn, very rapidly Landau damp and efficiently accelerate particles
to ultra-high energies ($\sim 0.1-1$PeV). The gravitational energy is thus converted to kinetic energy via 
electric field energy. This process requires initially accelerated electrons, which, by means of the frozen-in condition, will be supported by direct centrifugal mechanism. Charged electrons moving along co-rotating almost straight open magnetic field lines (see Fig. 1) will be energized, attaining large azimuthal velocity close to the light cylinder surface (a hypothetical zone,
where the linear velocity of rotation exactly equals speed of
light).

Searching for mechanism that  may boost particle energies to the PeV range
becomes highly relevant in the context of observations of the
Fermi-LAT collaboration - in particular,
 the observations of gamma-ray flares from the Crab
pulsar in $2009$ and $2010$ \citep{fermi_pev},  By analyzing the
spectral behaviour of high energy emission in the energy band
$>1$GeV, the authors conclude that electrons must have energies as
high as $1$PeV \citep{fermi_pev}.


The new contribution of this paper will consist, primarily, of exploring the possible role of the LLCD accelerated PeV
electrons in creating energetic photons. In section 2, we summarize the basic 
content of LLCD-the  basics of electron acceleration, and energy
loss/energy conversion mechanisms. In section 3, we work out  the
action of the accelerated electrons to create the PeV photons, the end
product of the processes explored by \cite{screp}. The Crab nebula
will be our representative millisecond pulsar. In section 4, we
summarize our findings and results.

\begin{figure}
 \centering
  \resizebox{5cm}{!}{\includegraphics[angle=0]{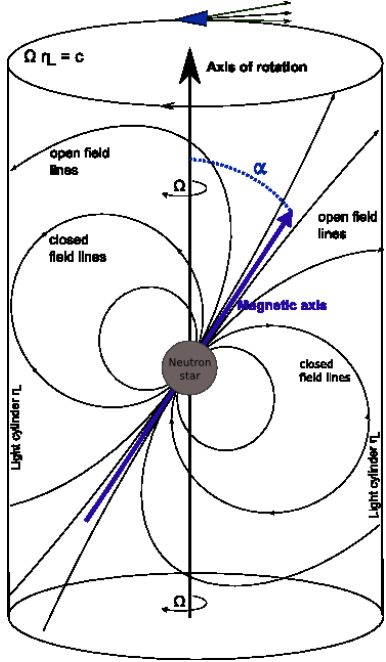}}
  \caption{Sketch of the magnetospheric-centrifugal Pulsar model. Charged particles
  sliding along co-rotating open magnetic field lines will attain a large azimuthal velocity component close to light cylinder surface.} \label{fig1}
\end{figure}

\begin{figure}
  \resizebox{\hsize}{!}{\includegraphics[angle=0]{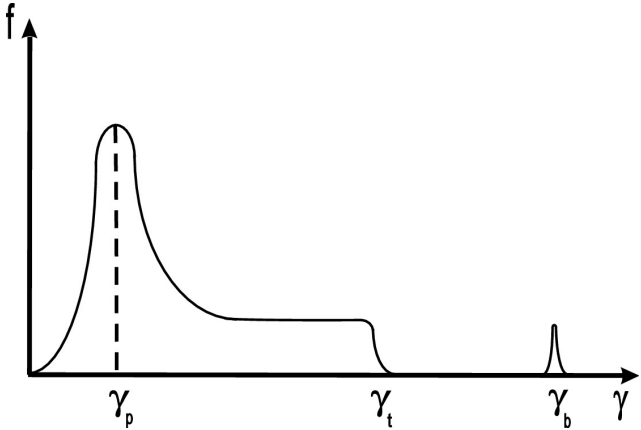}}
  \caption{The distribution function versus the Lorentz factor.
  As is clear from the plot, the function consists of two parts: the first -
  wider region concerns the plasma component corresponding
  to the cascade processes of pair creation and the second one characterizes the primary
  Goldreich-Julian beam electrons.}\label{fig2}
\end{figure}

\section[]{Theoretical model}

The major results of \cite{screp} were obtained in a simple
theoretical model in which the field lines are almost rectilinear
and the particle distribution corresponds to one in the standard
model of pulsars (Fig. \ref{fig2}). Here the narrower area
corresponds to the primary beam components and the wider one - to
the secondary particles. The linearized system of equations- the
Euler, the continuity, and the Poisson equation, governing the two
stream instability, then, reduces to the following form
\citep{screp,incr} ($e$ and $m$ are electron charge and rest mass) :
\begin{equation}
\label{eul} \frac{\partial p_{\beta}}{\partial
t}+ik\upsilon_{\beta0}p_{\beta}=
\upsilon_{\beta0}\Omega^2r_{\beta}p_{\beta}+\frac{2e}{m}E,
\end{equation}
\begin{equation}
\label{cont} \frac{\partial n_{\beta}}{\partial
t}+ik\upsilon_{{\beta}0}n_{\beta}, +
ikn_{{\beta}0}\upsilon_{\beta}=0
\end{equation}
\begin{equation}
\label{pois} ikE=4\pi e\sum_{\beta}n_{\beta0};
\end{equation}
where $\beta = 1,2$ is the species (electron, positron)index,
$p_{\beta}$ ($n_{\beta}$) is the perturbed momentum (the number
density), $\upsilon_{\beta}\approx V_{\beta}\cos\left(\Omega
t+\phi\right)$ and $n_{\beta0}$ are the unperturbed velocity and
number density, while
$r_{\beta}\approx\frac{V_{\beta}}{\Omega}\sin\left(\Omega
t+\phi\right)$ is the radial coordinate, and  $\Omega$ is the
angular velocity of rotation. the characteristic wave
number of the perturbed electrostatic field $E$
(of the induced Langmuir wave) is denoted by  $k$.
In what follows, we will recapitulate the salient features of the
content of \citep{galaxy,screp,incr}. 

The anzatz
\begin{equation}
\label{ansatz}
n_{\beta}=N_{\beta}e^{-\frac{iV_{\beta}k}{\Omega}\sin\left(\Omega t
+ \phi_{\beta}\right)},
\end{equation}
converts Eqs. (\ref{eul}-\ref{cont}) into a pair of coupled,
non-autonomous differential equations
\begin{equation}
\label{ME1} \frac{d^2N_1}{dt^2}+{\omega_1}^2 N_1= -{\omega_1}^2 N_2
e^{i \chi},
\end{equation}
\begin{equation}
\label{ME2} \frac{d^2N_2}{dt^2}+{\omega_2}^2 N_2= -{\omega_2}^2 N_1
e^{-i \chi},
\end{equation}
where $\chi = b\cos\left(\Omega t+\phi_{+}\right)$, $b = 2ck/\Omega
\sin\phi_{-}$, $2\phi_{\pm} = \phi_1\pm\phi_2$ and
$\omega_{1,2}\equiv\sqrt{8\pi e^2n_{1,2}/m\gamma_{1,2}^3}$ and
$\gamma_{1,2}$ are the relativistic plasma frequencies and the
Lorentz factors for the stream components. The two species have
different Lorentz factors and the corresponding centrifugal forces
are different as well. The centrifugal force, however, has a periodic time
dependentce, and, thus, capable of parametrically driving the
electrostatic waves.

A Fourier transform of Eqs. (\ref{ME1},\ref{ME2}) leads to the
"dispersion relation"
\begin{equation}
\label{disp} \omega^2 -\omega_1^2 - \omega_2^2  J_0^2(b)= \omega_2^2
\sum_{\mu} J_{\mu}^{2}(b) \frac{\omega^2}{(\omega-\mu\Omega)^2},
\end{equation}
where $J_{\mu}(x)$ is the Bessel function. The form of the preceding
expression suggests that  the system may undergo an instability at
the following resonance condition, $\omega_r = \mu_{res}\Omega$. In
the vicinity of the resonance,
 $\omega = \omega_r+\Delta$, the dispersion may be  approximated as
 \begin{equation}
 \label{disp1}
 \Delta^3=\frac{\omega_r {\omega_2}^2 {J_{\mu_{res}}(b)}^2}{2},
 \end{equation}
implying an imaginary part - the growth rate:
\begin{equation}
 \label{grow}
 \Gamma= \frac{\sqrt3}{2}\left (\frac{\omega_r {\omega_2}^2}{2}\right)^{\frac{1}{3}}
 {J_{\mu_{res}}(b)}^{\frac{2}{3}},
\end{equation}
where $\omega_r = \omega_1^2 + \omega_2^2  J_0^2(b)$

Through this centrifugally induced Langmuir instability, the
rotation energy is very efficiently pumped from the central engine -
the pulsar- to the electric field sustained by the bulk
magnetospheric e-p plasma. This is the first step in the
acceleration mechanism. In the second step, these waves Landau damp
on the much faster primary beam electrons, efficiently accelerating
them. The corresponding rate of generation of electrostatic waves is
given by \citep{vkm}
\begin{equation}
\label{dampr}  \Gamma_{LD}=
\frac{n\gamma\omega_p}{n_1\gamma_1^{5/2}},
\end{equation}
where $\omega_p=\sqrt{4\pi e^2 n/m}$, $\gamma$ and $n$ are the
plasma frequency, the Lorentz factor and the number density of the
specie on which the damping occurs.

The most optimum scenario for an overall efficient energy
pumping/transfer system is  realized when the instability growth and
Landau damping rates are large and comparable,
$\Gamma\sim\Gamma_{LD}$. If the instability growth rates were far in
excess of Landau damping rates, the waves will not be able to
transfer much energy to the particles. On the contrary, if the
Landau damping rates were much larger than the growth rates, the
waves will not grow much. In either extreme, there will be  very
little transfer from the star rotation to the particles.

For the crab pulsar parameters, it was been shown in \cite{screp} that the "optimizing" condition
$\Gamma\sim\Gamma_{LD}$ can be satisfied, for example, by the combination
$\gamma_2=4400$, $\gamma_1\approx 800$ corresponding to the so-called plasma component with moderate Lorentz factors. Here, $\gamma_1$ is the relativistic factor of electrons and $\gamma_2$ - the Lorentz factor of positrons. On the other hand, from symmetry it is clear that the similar results will be obtained if we interchange electrons and positrons. In addition,
the instability is supposed to be efficient in the
pulsar's magnetosphere if the corresponding timescale does not
exceed the pulsar's rotation period. This condition is well satisfied for
the aforementioned parameters. 

It is essential to probe  further into the details of energy
transfer from the star rotation into the plasma particles. In the
local frame of reference the particles are forced to slide along the
magnetic field lines by means of the centrifugal force. Viewed in
the laboratory frame, the reaction force
$F_{reac}\approx2mc\Omega\xi^{-3}$ \citep{grg}, where
$\xi\sim\gamma_1^{-1/2}$ \citep{rm00}, driving the particles,
becomes infinite on the light cylinder surface. This is a natural result- to preserve rigid rotation, the
particle velocity, in this region, must exactly equal the speed of
light. Note also that the radial velocity tends to zero, because on
the light cylinder  the particles can only rotate with the linear
velocity $c$. It is clear that the maximum energy gained from the
rotator can be estimated as the work done by the reaction force. For
all particles involved in the process, the energy gain may be
estimated as
\begin{equation}
\label{work}  W\approx n_1\delta VF_{reac}\delta r,
\end{equation}
where $\delta V$ is the volume in which the pumping takes place and
$\delta r\sim c/\Gamma$ is the corresponding lengthscale. This work
is transferred to the beam particles in the same volume, therefore
one should equate $W$ with $n_b\delta V\epsilon$ leading to the
estimated value of the maximum attainable energy of electrons
\begin{equation}
\label{en}  \epsilon\approx \frac{n_1F_{reac}\delta r}{n_b}.
\end{equation}
Simple estimates show that for  $\gamma_2=4400$, $\gamma_1\approx
800$ the primary electrons (beam components) with initial Lorentz factor, $10^7$, will
reach energy of the order of $\sim 100$TeV. To arrive at this number,
we have invoked  the Goldreich-Julian number density
$n_{_{GJ}} = \Omega(1-r^2/R_{lc}^2)^{-1}/ B/2\pi ec$ \citep{rud},
where $B = B_{st}(R_{st}/R_{lc})^3$ ($B_{st}\approx 6.7\times
10^{12}$ is the magnetic field close to the neutron star's surface
and $R_{lc}\equiv cP/(2\pi)$ is the light cylinder radius). It is also
assumed that energy is uniformly distributed among various magnetospheric
"species" of particles, $n_1\gamma_1\approx n_2\gamma_2\approx
n_{_{GJ}}\gamma_{_{GJ}}$.

When we invoke another combination $\gamma_2=2\times 10^4$,
$\gamma_1\approx 4\times 10^3$ that satisfies the condition
$\Gamma\sim\Gamma_{LD}$, and for which the instability time scale
$1/\Gamma$ does not exceed the kinematic timescale, $P$ (insuring
efficient energy  transfer), the projected electron energy jumps up
to $\sim 1.3$PeV. Thus, the centrifugally excited Langmuir waves can
readily accelerate electrons to PeV energies in the magnetosphere of
the crab nebula.

\section{Generation of PeV photons}

What phenomena will follow in the wake of ultra high energy (PeV) particles interacting with the strong magnetic fields in the pulsar magnetosphere? Copious synchrotron losses will result-the synchrotron emission,
\begin{equation}
\label{Psyn}  P_{syn}\approx \frac{2e^4\epsilon^2B^2}{3m^4c^7},
\end{equation}
which is so strong that the cooling (energy-loss) time $t_{syn}\sim
\epsilon/P_{syn}\sim 10^{-13}$s  turns out to be much smaller than
the kinematic timescales.

Notice that the aforementioned classical expression for power
emission must be appropriately corrected when quantum effects become
important. The quantum modification of the synchrotron mechanism
is controlled by the parameter \citep{sokolov}
\begin{equation}
\label{xi}  \xi\equiv \frac{3}{2}\gamma \overline{B}\sin\psi,
\end{equation}
where $\gamma\equiv\epsilon/mc^2$ is the Lorentz factor of the ultra
relativistic electrons, $\overline{B}\equiv B/B_{cr}$,
$B_{cr}=m^2c^3/(\hbar e)\approx 4.4\times 10^{13}$G is the Schwinger
limit for the magnetic induction and $\psi$ is the pitch angle of
particles. For particles with $\epsilon\sim 1$PeV, the
aforementioned parameter is of order $10^2$ implying that we are
deep in the quantum domain. In \cite{sokolov} the authors argue that the
total energy emitted by electrons in the ultra
quantum case may be  approximated  by
\begin{equation}
\label{Pq}  P_{q} =
\frac{2^{8/3}}{9}\frac{\Gamma\left(\frac{2}{3}\right)}{\xi^{-4/3}}P_{cl},
\end{equation}
where $\Gamma(x)$ is the  Euler Gamma function. Even in  the quantum
domain, the estimated  cooling time scale is still extremely small,
and therefore, the  quantum synchrotron radiation remains an
efficient energy loss mechanism. The power spectrum of "quantum"
synchrotron emission of a single particle is given by
\citep{brainerd}
\begin{equation}
\label{w}  \frac{dW}{d\omega dt} =
\frac{\sqrt{3}}{2\pi}\frac{e^2mc\overline{B}\sin\psi}{\hbar}F,
\end{equation}
where
\begin{equation}
\label{F}  F_{\xi}\left(x\right) =
x\int_{x/(1-x\xi)}^{\infty}K_{5/3}(x')dx'+\frac{x^3\xi^2}{1-x\xi}K_{2/3}\left(\frac{x}{1-x\xi}\right),
\end{equation}

\begin{equation}
\label{x}  x =
\frac{2}{3}\frac{\epsilon_{ph}}{\gamma^2\overline{B}\sin\psi},
\end{equation}
and $\epsilon_{ph}$ is energy of the synchrotron photon in units of
$mc^2$.  The function $F_{\xi}(x)$ reduces to the classical
expression when the parameter $\xi$ goes to zero.

In the opposite limit, $\xi \gg 1$, deviations from the classical 
expression become highly significant. In fact,  "discrete" quantum processes
become dominant and  the radiation process ceases to be continuous. 
In this domain of primary interest (to this paper), the standard process of synchrotron self Compton (SSC) mechanism 
may not be the most effective in determining the very high energy emission of pulsars because
SSC requires that the relativistic particles continuously
emit synchrotron photons. The ultra relativistic particles,
however, will likely lose almost their entire kinetic energy in a
single quantum event - the emission of a very energetic photon and
as a result the emission is not continuous. Consequently, the SSC,
requiring a sea of photons (in turn, continuously emitted by the
same particle), becomes  much less probable.

For electrons with PeV energies, $\xi\sim100$ with the corresponding  $x\sim.01$ (at maximum
synchrotron emission) \citep{brainerd}. Equation (\ref{x}), then, tells us that the
radiated photons will have energies
of the order of $\sim 0.6$PeV. Thus the ultra high energy electrons
rapidly produce ultra-high energy photons before they leave the star's magnetosphere. In fact, 
before exit from the pulsar vicinity, they might lose a large part of their energy.

The next step, therefore, is to investigate how transparent the
medium of the Crab nebula is to  these ultra-high energy gamma-rays.
The principal modes controlling the propagation of these high energy
photons will be : 1) interaction  with other but  softer photons, 2)
interaction with nebular matter.

Photon - photon interaction leads to efficient pair creation when
the soft photons have frequency of the order  \citep{aharonian}
\begin{equation}
\label{gg}  \nu(Hz)\approx4\frac{m^2c^4}{\epsilon_{ph}}\approx
2.4\times 10^{11}\frac{PeV}{\epsilon_{ph}}.
\end{equation}
The corresponding  maximum cross section is approximately
\citep{aharonian}
\begin{equation}
\label{sgg}  \sigma_{\gamma\gamma}\approx 0.2\sigma_T.
\end{equation}
Equation (\ref{gg}) reveals that the $\gamma\gamma$ interaction of
the  $\sim 0.1$PeV - $1$PeV photons will be highly efficient  with
very low energy soft photons. From \citep{hester} it is clear that
for photons with $\nu\sim 10^{11}$Hz, the luminosity of the Crab
nebula is of the order of $\sim 10^{35}$erg/s. This in turn means
that the number density of these photons,
\begin{equation}
\label{ngg}  n_{\gamma\gamma}\approx\frac{L}{4\pi r_n^2ch\nu}\approx
3\times\frac{PeV}{\epsilon_{ph}}cm^{-3},
\end{equation}
is so small that the lengthscale of the $\gamma\gamma$ interaction
\begin{equation}
\label{lgg}
\lambda_{\gamma\gamma}\sim\frac{1}{\sigma_{\gamma\gamma}n_{\gamma\gamma}}\approx
2.5\times 10^{24}\frac{\epsilon_{ph}}{PeV}cm,
\end{equation}
exceeds the Crab nebula size by several orders of magnitude.
Therefore, this particular part of interaction may not be very
important and we neglect it.

The ultra high energy photons
with $\sim 0,6$PeV might encounter also the $\nu\approx 3\times 10^{11}$Hz
cosmic microwave background (CMB) soft photons. The problem of
absorption of ultra-high energy photons by CMB photons was
considered in detail by \cite{kohri}. The authors also plotted
dependence of attenuation length on photon's energy. In our case,
the corresponding value is $\lambda\simeq 50$kpc, which means that the
absorption factor, $f$, is given by
\begin{equation}
\label{abs}
f\simeq \exp{\left(-d/\lambda\right)}\simeq 0.96.
\end{equation}
Therefore, a very small fraction, only $4\%$ of $0.6$PeV photons which potentially could
have been reached the earth, will undergo efficient pair creation
producing electrons, with energies 
\citep{aharonian}
\begin{equation}
\label{rang}
\frac{\epsilon_{ph}}{2}\left(1-\sqrt{1-\frac{m^2c^4}{h\nu\epsilon_{ph}}}\right)
\leq\epsilon_e\leq\frac{\epsilon_{ph}}{2}\left(1+\sqrt{1-\frac{m^2c^4}{h\nu\epsilon_{ph}}}\right).
\end{equation}
in the following interval $40$TeV and $600$TeV respectively.

By taking into account that pulsars emit in two relatively narrow channels, one can straightforwardly arrive at an approximate value of the net flux of $600$ TeV energy photons
\begin{equation}
\label{flux}
\frac{dN}{dAdt}\simeq 2\;\alpha\;c\;n_{GJ}\times\left(\frac{R_{lc}}{d}\right)^2,
\end{equation}
where $\alpha$ is a fraction of magnetospheric electrons involved in the acceleration process up to $600$ TeV energies and $d\simeq 2$kp is the distance from the Crab pulsar.

In \citep{borione} the authors examined $2.4\times 10^9$ events detected by the Chicago Air Shower Array-Michigan Muon Array (CASA-MIA) experiment to study ultra-high energy ($>100$TeV) gamma-rays from the Crab pulsar and nebula. It has been shown that an integral flux limit from the Crab nebula is of the order of $1.63\times 10^{-16}$ cm$^{-2}$s$^{-1}$. After combining this value with Eq. (\ref{flux}) one can show that $\alpha\simeq 4\times 10^{-7}$.

The derived number is a significant value because it indicates the fraction of particles involved in the generation of ultra-high energies. On the other hand, it is strongly believed that pulsars might contribute to the generation of cosmic rays, \citep{hooper} and therefore, in studying the population of very high energy cosmic particles, the estimated parameter is significant. It is worth noting that $\alpha$ might be an indirect indicator of the topology of the magnetic field of the pulsar's magnetosphere and, sooner or later, we are going to study this particular problem as well.

\section{Summary}

     We have explored the consequences of a mechanism of electron's
      magneto-centrifugal acceleration and applied it to the Crab pulsar. According to
      this mechanism, the energy of  the rotator (pulsar) is efficiently
      pumped to the magnetospheric plasma exciting Langmuir
      waves, which, efficiently Landau damp on the primary electrons
      accelerating them to enormous energies of the order of $\sim
      1$PeV.

      We show that these $\sim 1$PeV particles efficiently radiate
      in the quantum synchrotron regime, producing photons with $\sim 0.6$PeV.
      These ultra-high energy gamma
      rays travel through the nebula  to the interstellar medium without  much energy loss.
      In the interstellar region, however, these photons encounter
      the CMB soft photons, and only small portion of them ($4\%$) by the $\gamma\gamma$
      channel efficiently produce very energetic electron-positron pairs.

      By considering the observed integral flux of $600$ TeV gamma-rays
      and comparing with the model we have estimated the fraction of magnetospheric particles 
      involved in generation of PeV energy electrons, $\alpha\simeq 4\times 10^{-7}$.


\section*{Data availability statement}
All data generated or analysed during this study are included in this published article.
\bibliographystyle{spr-mp-nameyear-cnd}

\begin{thebibliography}{99}

\bibitem[\protect\citeauthoryear{Osmanov}{2021}]{galaxy} Osmanov, Z.N., Relativistic Effects of Rotation in $\gamma$-ray Pulsars?Invited Review. Galaxy 2021; 9:1-17.
\bibitem[\protect\citeauthoryear{Mahajan et al.}{2013}]{screp}
Mahajan, S., Machabeli, G., Osmanov, Z. \& Chkheidze, N., Ultra High Energy Electrons Powered by Pulsar Rotation. Nat. Sci. Rep. 2013;  3:1-5.
\bibitem[\protect\citeauthoryear{Machabeli et al.}{2005}]{incr} Machabeli, G.,
Osmanov Z. \& Mahajan, S., Parametric mechanism of the rotation energy pumping by a relativistic plasma. Phys. Plasmas 2005; 12:1-12.
\bibitem[\protect\citeauthoryear{Deutsch}{1955}]{deutsh} Deutsch, A., The electromagnetic field of an idealized star in rigid rotation in vacuo. Ann. Astrophys. 1955;  18:1-10.
\bibitem[\protect\citeauthoryear{Goldreich \& Julian}{1969}]{gj} Goldreich, P. \& Julian, W.H., Pulsar Electrodynamics. ApJ 1969; 157:869-880.
\bibitem[\protect\citeauthoryear{Abdo et al.}{2011}]{fermi_pev} Abdo, A.A. et al., Gamma-Ray Flares from the Crab Nebula. Science 2011; 331:739-742.
\bibitem[\protect\citeauthoryear{Volokitin et al.}{1987}]{vkm} Volokitin, A.S., Krasnoselskikh, V.V. \& Machabeli, G.Z., Waves in the relativistic electron-positron plasma of a pulsar. Sov. J. of Pl. Phys.  1987; 11: 310-314.
\bibitem[\protect\citeauthoryear{Rogava et al.}{2003}]{grg} Rogava, A., Dalakishvili, G. \&
Osmanov, Z., Centrifugally Driven Relativistic Dynamics on Curved Trajectories. Gen. Rel. Grav. 2003; 35:1133-1152.
\bibitem[\protect\citeauthoryear{Rieger \& Mannheim}{2000}]{rm00} Rieger, F.M. \& Mannheim, K., Particle acceleration by rotating magnetospheres in active galactic nuclei.  A\&A 2000; 353:473-478.
\bibitem[\protect\citeauthoryear{Ruderman \& Sutherland}{1975}]{rud} Ruderman, M.A. \& Sutherland, P.G., Theory of pulsars: polar gaps, sparks, and coherent microwave radiation.  ApJ 1975; 196:51-72.
\bibitem[\protect\citeauthoryear{Sokolov \& Ternov}{1968}]{sokolov} Sokolov, A.A. \&
Ternov, I.M., Synchrotron Radiation. New York: Pergamon; 1968.
\bibitem[\protect\citeauthoryear{Brainerd}{1987}]{brainerd} Brainerd, J.J., Quantum Synchrotron Spectra from Semirelativistic Electrons in Teragauss Magnetic Fields. ApJ 1987; 320:714-725.
\bibitem[\protect\citeauthoryear{Aharonian}{2004}]{aharonian} Aharonian, F.A., Very High Energy Cosmic Gamma Radiation - A Crucial Window on the Extreme Universe. World Scientific Publishing Co.
Pte. Ltd.; 2004.
\bibitem[\protect\citeauthoryear{Hester}{2008}]{hester} Hester, J.
Jeff, The Crab Nebula : an astrophysical chimera. ARA\&A  2008; 46:127-155.
\bibitem[\protect\citeauthoryear{Kohri et al.}{2012}]{kohri} Kohri, K., Ohira, Y. \& Ioka, K., Gamma-ray flare and absorption in the Crab nebula: lovely TeV-PeV astrophysics. MNRAS 2012; 424:2249-2254.
\bibitem[\protect\citeauthoryear{Borione et al.}{1997}]{borione} Borione, A. et al., A Search for Ultra-High-Energy Gamma-Ray Emission from the Crab Nebula and Pulsar.  ApJ 1997; 481:313--326.
\bibitem[\protect\citeauthoryear{Hooper et al.}{2009}]{hooper} Hooper, D., Blasi, P. \& Serpico, P.D., Pulsars as the sources of high energy cosmic ray positrons. JCAP 2009; 1:1--13.


\end{thebibliography}

\end{document}